\documentstyle[11pt]{amsart}

\begin{document}

\title[Hyperbolic Topological Invariants and the Black Hole Geometry]
{Hyperbolic Topological Invariants and the Black Hole Geometry}

\author{A. A. Bytsenko}
\address{Departamento de F\'isica, Universidade Estadual de Londrina,
Caixa Postal 6001, Londrina--Parana, Brazil \,\,
{\em E-mail address}: {\rm abyts\, at \,uel.br}}

\author{E. Elizalde}
\address{Institut d'Estudis Espacials de Catalunya (IEEC/CSIC),
Edifici Nexus, Gran Capit\`{a} 2--4, 08034 Barcelona,
Department d'Estructura i Constituents de la Mat\`{e}ria
Facultat de F\'{\i}sica, Universitat de Barcelona,
Av. Diagonal 647, 08028 Barcelona, Spain \,\,
{\em E-mail address:} {\rm elizalde\, at \,ieec.fcr.es}}

\author{S. A. Sukhanov}
\address{State University of Sankt Petersburg,
199034 Universitetskaja nab. 7--9, Sankt Petersburg, Russia}

\date{February, 2003}

\thanks{
The research of A.A.B. has been supported in part by the Russian
Foundation for Basic Research (grant No. 01--02--17157). E.E. has
been supported by DGI/SGPI (Spain), project BFM2000-0810, and by
CIRIT (Generalitat de Catalunya), contract 1999SGR-00257.}

\maketitle

\begin{abstract}

We discuss the isometry group structure of three-dimensional black
holes and Chern--Simons invariants. Aspects of the holographic
principle relevant to black hole geometry are analyzed.

\end{abstract}

\section{Introduction}

The discovery of black hole solutions in three--dimensional gravity offers
a promising new area for the analysis of interesting and difficult problems
that were posed in the four-dimensional case.
We begin here with the discussion of some
geometrical aspects of the non--rotating, three--dimensional black hole
\cite{Banados}. It is known that a static Lorentzian metric  is a
solution
of the three--dimensional vacuum Einstein equation with negative cosmological
constant $\Lambda$,
i.e. $R_{\mu \nu}=2\Lambda g_{\mu \nu}$\,,\,\, $R=6\Lambda=- 6\sigma^{-2}$.
The sectional curvature $k$ is constant and negative, namely
$k= - \sigma^{-2}$ .
The metric describes a space--time which is locally isometric to the anti--de
Sitter space.
The Euclidean section is obtained by the Wick rotation $t\rightarrow i\tau$,
and reads
$$
ds^2=\left(r^2 \sigma^{-2}-M \right) d\tau^2+\left(r^2 \sigma^{-2}-M
\right)^{-1}dr^2+r^2 d\varphi^2
\:,
\eqno{(1.1)}
$$
where the coordinates $(t,r,\varphi)$ have been used (8G=1 is assumed,
so that the mass $M$ is dimensionless),
and $\sigma$ is a dimensional constant.
The metric (1.1) has a horizon radius given by $r_+=\sigma M^{1/2}$.
With a change of coordinates, $(\tau,r,\varphi) \to (y, x_1,x_2)$, of the form
$$
y =\frac{r_+}{r}\exp\left({\frac{r_+}{\sigma}\varphi}\right)\:,
\,\,\,\,\,\,\,\,\,\,
x_1+ix_2 = \frac{1}{r} \left(r^2-r_+^2\right)^{1/2}
\exp\left(i\frac{r_+}{\sigma^2}\tau
+\frac{r_+}{\sigma}\varphi\right) \:,
\eqno{(1.2)}
$$
the metric becomes
$$
ds^2=\frac{\sigma^2}{y^2}\left( dy^2+dx_1^2+dx_2^2\right)
\:.
\eqno{(1.3)}
$$
As a consequence, the metric
describes a manifold homeomorphic to the real hyperbolic space ${\Bbb H}^3$.
The group of isometries of ${\Bbb H}^3$ is  $SL(2,\Bbb C)$.

Let us now consider a discrete subgroup $\Gamma\subset
PSL(2,{\Bbb C})\equiv SL(2,{\Bbb C})/\{\pm Id\}$ ($Id$ is the
identity element), which acts discontinuously at the point $z$
belonging to the extended complex plane ${\Bbb
C}\bigcup\{\infty\}$. We recall that a transformation $\gamma\neq
Id$, $\gamma\in\Gamma$, with $\gamma z=(az+b)(cz+d)^{-1}$\,,\,
$ad-bc=1$\,, \, $a,b,c,d\in {\Bbb C}$\,, is called elliptic if
$({\rm Tr}\gamma)^2=(a+d)^2$ satisfies $0\leq({\rm
Tr}\gamma)^2<4$, hyperbolic if $({\rm Tr}\gamma)^2>4$, parabolic
if $({\rm Tr}\gamma)^2=4$, and loxodromic if $({\rm
Tr}\gamma)^2\in {\Bbb C}\backslash [0,4]$. The periodicity of the
angular coordinate $\varphi$ allows to describe the black hole
manifold as the quotient $\Gamma\backslash  {\Bbb H}^3$, $\Gamma$
being a discrete group of isometries possessing a primitive
element $\gamma_h \in \Gamma$, defined by the identification
$\gamma_h(y,w)=(\exp (2 \pi r_+ \sigma^{-1}) y$,\,\, $\exp (2 \pi
r_+ \sigma^{-1}w))\sim (y,w)$. Therefore, the matrix
$$
\gamma_h= \left(
\begin{array}{cc}
e^{\frac{r_+}{\sigma}} & 0 \\
0 & e^{-\frac{r_+}{\sigma}}
\end{array}
\right)
\:,
\eqno{(1.4)}
$$
corresponds to an hyperbolic element (${\rm Tr} \gamma_h >2$)
consisting in a pure
dilatation. Furthermore, since in  Euclidean  coordinates $\tau$
becomes an angular variable, with period $\beta$, one is led also to the
identification
$\gamma_e(y,w)=(y,\,\,\exp (i\beta r_+ \sigma^{-2})w)\sim (y,w)$.
This identification is generated by an elliptic element in the group $\Gamma$,
namely
$$
\gamma_e= \left(
\begin{array}{cc}
e^{i\beta \frac{r_+}{\sigma^2}} & 0 \\
0 & e^{-i\beta \frac{r_+}{\sigma^2}}
\end{array}
\right)
\:,
\eqno{(1.5)}
$$
as long as ${\rm Tr} \gamma_e <2$, and a conical singularity will be
present. If $\beta r_+ \sigma^{-2}=2 \pi$,
then $\gamma_e=Id$ and the conical singularity is absent. As a result,
the period is determined to be
$\beta_H=2 \pi\sigma^2(r_+)^{-1}$
and this is interpreted as the inverse of the Hawking temperature.
The $M=0$ line element has the form
$ds^2_0= r^2\sigma^{-2}d\tau^2+r^{-2}\sigma^2 dr^2+r^2d\varphi^2$.
It is clear that $\tau$ can be identified with
any period $\beta$ (in particular with $\beta=\infty$), and that $\varphi$
has period
$2\pi$. Through the coordinate change:  $r=y^{-1}\sigma^2$,\, $\tau=x_1$,
 $\varphi=x_2\sigma^{-1}$, one gets  the metric of the hyperbolic space:
$ds^2_0=\sigma^2 y^{-2} (dx_1^2+dx_2^2+dy^2)$, and the identification
$\gamma_p(w,y)=(w+\beta + 2i\pi\sigma\,y) \simeq (w,y)$.
This identification is generated by elements of $\Gamma$ of the form
$$
\gamma_{p_1}= \left(
\begin{array}{cc}
1 & \beta \\
0 & 1
\end{array}
\right) \,\,,\,\,\,\,\,\,\,\,\,\,\,
\gamma_{p_2}= \left(
\begin{array}{cc}
1 & 2i\pi  \sigma \\
0 & 1
\end{array}
\right)
\:,
\eqno{(1.6)}
$$
which are parabolic. Therefore the on--shell three--dimensional
black hole can be regarded as a strictly hyperbolic, non--compact
manifold $\Gamma\backslash{\Bbb H}^3$ (see Sect. 2).

The group $SL(2, {\Bbb C})$ being the universal covering of the Lorentz group,
for a Euclidean space there are
one--forms which can be combined to build a single $SL(2, {\Bbb C})$
connection $A^{a}_{\mu}$. It has been shown \cite{Witten} that the
 Einstein action
can be reduced to the Chern--Simons action associated with the
connection $A^{a}_{\mu}$. $SL(2, {\Bbb C})$ gauge transformations
in the Chern--Simons formulation are equivalent to diffeomorphisms
and local Lorentz transformations in the metric formulation of the
black hole \cite{Carlip}. In Sect. 3 we will work with the
Chern--Simons formulation, which follows from the relationship
(non-trivial, in fact, since it has a nonvanishing
holonomy) between flat connections
 and geometric structures related to
one--forms. Some aspects of the holographic principle for hyperbolic geometry
will be considered in Sect. 4.

\section{The Arithmetic Geometry of 
$\Gamma = SL(2,{\Bbb Z}+i{\Bbb Z})/\{\pm Id\}$}

Here we shall summarize the geometry and local isometry associated
with a simple three--dimensional complex Lie group. We shall
consider that the discrete subgroup $\Gamma$ acts discontinuously
at the points of the extended complex plane. The isometric circle
of a transformation $\gamma\in G = SL(2,{\Bbb C})/\{\pm Id\}$ for
which $\infty$ is not a fixed point is defined to be the circle
$I(\gamma)=\{z: \,\,|\gamma z|=1\}$, or $I(\gamma)= \{z: \,\,
|z+d/c|=|c|^{-1}\}$,\, $c\neq 0$. A transformation $\gamma$
carries its isometric circle $I(\gamma)$ into $I(\gamma^{-1})$.
For $\Gamma\subset G$, one can choose a subgroup $\Gamma$ of the
form $SL(2,{\Bbb Z}+i{\Bbb Z})/\{\pm Id\}$, where ${\Bbb Z}$ is
the ring of integers. The element $\gamma\in\Gamma$ will be
identified with $-\gamma$. The group $\Gamma$ has, within a
conjugation, one maximal parabolic subgroup $\Gamma_\infty$. The
fundamental domain related to $\Gamma$ has one parabolic vertex,
and can be taken to be of the form: $F(\Gamma)=\{ (y,w): x_1^2+
x_2^2 +y^2
>1,\:\:\: -1/2 < x_2 <x_1< 1/2 \}$.

The Selberg trace formula can be constructed as an expansion in
eigenfunctions of the automorphic Laplacian. Since the discrete
group $\Gamma$ has a cusp at $\infty$, each element of the
stabilizer $\Gamma_{\infty}$ is a translation. The conjugacy class
$\{\gamma\}_{\Gamma}$, $\gamma\in\Gamma_{\infty}$, with $\gamma$
different from the identity, and the centralizers related to this
class of $\gamma$, have been calculated in Refs. 4,5. 
Let $\lambda_j$ be the isolated
eigenvalues of the self--adjoint extension of the Laplace operator
and let us introduce a suitable analytic function $h(r)$ and
$r^2_j=\lambda_j-1$. For one parabolic vertex  let us introduce a
subdomain $F_{\mathcal Y}$ of the fundamental region $F(\Gamma)$
by means of $F_{\mathcal Y}=\{ z \in F(\Gamma),\, z=\{y, {\bf x}
\}|y\leq {\mathcal Y}\}$, where ${\mathcal Y}$ is a sufficiently
large positive number.

\medskip
\medskip

\par \noindent
{\bf Lemma 1}\,\,\, (Ref. [4]\,, Eq. (13)). \,\,\,\,\,
{\em Suppose $h(r)$ to be an even analytic function in
the strip $|\Im r|<1+\varepsilon $ ($\varepsilon>0$), and
$h(r)={\mathcal O}(1+|r|^2)^{-2}$.
Then the following formula holds:
$$
\sum_j h(r_j)=\lim_{{\mathcal Y}\rightarrow\infty}\left\{
\int_{F_{\mathcal Y}}\sum_{\{\gamma\}_{\Gamma}}k(u(z,\gamma z))\:d\mu(z)
\right.
$$
$$
\left.
-\frac{1}{2\pi}\int_{{\Bbb R}_+}h(r)
\int_{F_{\mathcal Y}}|E(z,1+ir)|^2\:d\mu(z)dr
\right\}
\mbox{,}
\eqno{(2.1)}
$$
where $d\mu(z)=y^{-3}dydx_1dx_2$ is the invariant measure on ${\Bbb H}^3$,
$k(z, z') = k(u(z, z'))$ is the kernel of the invariant operator,
$u(z, z')=|z-z'|^2/yy'$,
and $E(z,s)$ is the Eisenstein--Maass series associated with one cusp.
}
\\
\\
The final result (see for detail Ref. [5]) could be considered
as an addition to this Lemma 1.

\medskip
\medskip

\par \noindent
{\bf Theorem 2}\,\,\,
{\em For the special discrete group $SL(2,{\Bbb Z}+i{\Bbb Z})/\{\pm
Id\}$ and $h(r)$ satisfying the conditions of Lemma 1, the
Selberg's trace formula holds
$$
\sum_j h(r_j)-\sum_{\scriptstyle\{\gamma\}_{\Gamma},\gamma\not=Id,
\atop\scriptstyle\gamma-non-parabolic}\int k(u(z,\gamma z))\:d\mu(z)
= {\rm vol}(F)\int_{{\Bbb R}_+}\frac{r^2}{2\pi^2}\:h(r)\:dr
$$
$$
+ \frac{1}{4\pi}\int_{\Bbb R}h(r)
\left(\frac{d}{ds}{\rm log}\,S(s)|_{(s=1+ir)}
-\psi(1+ir/2)\right)
+\frac{h(0)}{4}\left(1 - S(1)\right)
+ C{\rm g}(0)
\mbox{.}
\eqno{(2.2)}
$$
Here ${\rm vol}(F)$ is the (finite) volume of the fundamental
domain $F$ with respect to the measure $d\mu$, the function $S(s)$
(in the general case it is the S--matrix) is given by a
generalized Dirichlet series, convergent for $\Re s>1$. The
functions $E(z,s)$ and $S(s)$ can be analytically extended on the
whole complex $s-$ plane, where they satisfy the functional
equations $S(s)S(1-s)=Id$,\,\,\, $\overline{S(s)}=S(\bar s)$,\,\,
$E(z,s)=S(s)E(z,1-s)$. $\psi (s)$ is the logarithmic derivative of
the Euler $\Gamma-$ function, $C$ a computable constant and ${\rm
g}(u)$ denotes the Fourier transform of $h(r)$.}

\subsection{The functional determinant}

In general, the determinant of an elliptic differential operator
requires a regularization. It is convenient to introduce the
operator ${\frak L}_{\Gamma}(\delta)= {\frak L}_{\Gamma} +
\delta^2-1$, with $\delta$ a suitable parameter. One of the most
widely used regularization is the zeta--function regularization
(see Refs. [6,7,8]). Thus one has ${\rm
log} {\rm det} {\frak L}_{\Gamma}(\delta)=-(d/ds)\zeta ( s|{\frak
L}_{\Gamma}(\delta))|_{(s=0)}$. In standard cases, the zeta
function at $s=0$ is well defined and  one gets a finite result.
The meromorphic structure of the analytically continued zeta
function is related to the asymptotic properties of the
heat--kernel trace. For the rank one symmetric space
$\Gamma\backslash {\Bbb H}^3$ the trace of the operator
$\exp[-(t{\frak L}_{\Gamma}(\delta))]$ could be computed by using
Theorem 2 (Eq. (2.2)) with the choice
$h(r)=\exp[-t(r^2+\delta^2)]$ (we use units in which the curvature
of ${\Bbb H}^3$ is equal to $-1$). Thus we have ${\rm g}(u)= (4\pi
t)^{-1/2} \exp(-t\delta^2 - u^2(4t)^{-1})$, \, ${\rm g}(0)= (4\pi
t)^{-1/2} \exp(-t\delta^2)$,\, and $h(0)=\exp(-t \delta^2)$.

We shall consider additive terms of the zeta function associated
with the identity and parabolic elements of the group $\Gamma$
only (the heat kernel and the zeta--function analysis for
co--compact discrete group $\Gamma$ has been done, for example, in
Refs. [6,7]),
$$
{\rm Tr}\left( \exp (-t ({\frak L}_{\Gamma}(\delta)-\delta^2))\right) =
\frac{{\rm vol}(F)+ 4\pi t C}{(4\pi t)^{3/2}}
+\frac1{4}-\frac{1}{4\pi}\int_{\Bbb R}
\psi(1+ir/2)e^{-tr^2}\:dr
\mbox{.}
\eqno{(2.3)}
$$
Making use of the trace formula, we compute the functional
determinant of the Laplace--type operator on $\Gamma\backslash
{\Bbb H}^3$. The zeta function for $\Re s$ sufficiently large, can
be rewritten in the form
$$
\zeta(s|{\frak L}_{\Gamma}(\delta))=\sum_\sigma\rho_\sigma\left
(\lambda_\sigma+\delta^2-1\right)^{-s}
=\sum_j \left( \lambda_j+\delta^2-1\right)^{-s}
$$
$$
+\int_{{\Bbb R}_+} \left( \lambda+\delta^2-1 \right)^{-s}\:
\rho_\lambda\:d\lambda
\mbox{,}
\eqno{(2.4)}
$$
where the sum over $j$ runs over the discrete spectrum,
$\{\lambda_j\}\in {\rm Spec}({\frak L}_{\Gamma})$. For the
continuous spectrum, $\rho_\lambda$ is proportional to the
logarithmic derivative of the S--matrix $S(s)$. 

One has
$(d/ds)\zeta (s|{\frak L}_{\Gamma}(\delta)) =-\sum_\sigma
\rho_\sigma \left(\lambda_\sigma+\delta^2-1\right)^{-s}{\rm
log}(\lambda_\sigma+\delta^2-1)$, and
$$
\frac{d}{d \delta} \left( \frac{1}{2\delta}\frac{d}{d \delta}
\frac{d}{ds}\zeta (s|{\frak L}_{\Gamma}(\delta))
\right)=2\delta
\sum_\sigma \rho_\sigma  \left( \lambda_\sigma+\delta^2-1 \right)^{-s-2}
+ {\mathcal O}(s)
\mbox{.}
\eqno{(2.5)}
$$
A standard Tauberian argument gives a Weyl's estimate for large
$\sigma$, namely \cite{Bytsenko}:
$(\lambda_\sigma+\delta^2-1)^{-1}\simeq \sigma^{-2/3}$. As a
consequence, in the limit $s \rightarrow 0 $, the right hand side
of Eq.~(2.5) is finite. This works for two-- and
three--dimensional cases. In higher dimensions it is necessary to
take further derivatives with respect to $\delta$ \cite{Voros}.
The inclusion of the contribution related to the hyperbolic
elements is almost straightforward and can be found, for example,
in Refs. [6,7]. It is additive and reads
simply ${\rm log} Z_{\Gamma}(1+\delta)$, like before
$Z_{\Gamma}(s)$ is the Selberg zeta function. Summarizing, the
final result ist:

\medskip
\medskip

\par \noindent
{\bf Theorem 3}\,\,\, (Ref. [5], Theorem 2). \,\,\,\,\,
{\em The following identity holds
$$
\det {\frak L}_{\Gamma}(\delta) = \frac{2}{(\pi\delta)^{1/2}\Gamma(\delta/2)}
\exp\left(-\frac{{\rm vol}(F)\delta^3}{6\pi}+C\delta\right)\,
Z_{\Gamma}(1+\delta)
\mbox{.}
\eqno{(2.6)}
$$
}

\section{Topological Invariants of the Hyperbolic Geometry}

The Chern--Simons partition function may be expressed through the
asymptotics which lead to a series of $C^{\infty}-$ invariants
associated with triplets $\{X;{\mathcal F};\xi\}$, with $X$ a
smooth homology three-sphere, ${\mathcal F}$ a homology class of
framings of $X$, and $\xi$ an acyclic conjugacy class of
orthogonal representations of the fundamental group $\pi_1(X)$
\cite{Axelrod}. In addition, the cohomology $H(X;{\rm Ad}\,\xi)$
of $X$ with respect to the local system related to ${\rm Ad}\,\xi$
vanishes.

We turn into Chern--Simons invariants related to real hyperbolic
spaces. Let ${\frak M}=G/K$ be an irreducible rank one symmetric
space of non--compact type, where $G$ is a connected non--compact
simple split rank one Lie group with finite centre, and $K\subset
G$  a maximal compact subgroup \cite{Helgason}. Let $\Gamma\subset
G$ be a discrete, co--compact torsion free subgroup. Then
$X=\Gamma\backslash {\frak M}$ is a compact Riemannian manifold
with fundamental group $\Gamma$, i.e. $X$ is a compact locally
symmetric space. For a real hyperbolic manifold, we have
$G=SO_1(n,1)$ \,$(n\in {\Bbb Z}_{+})$, \, $K=SO(n)$, and $X=\Gamma
\backslash {\Bbb H}^n$. Given a finite--dimensional unitary
representation $\chi$ of $\Gamma$, there is the corresponding
vector bundle $V_{\chi}\rightarrow X$ over $X$, given by
$V_{\chi}=\Gamma\backslash ({\frak M}\otimes F_{\chi})$, where
$F_{\chi}$ (the fibre of $V_{\chi}$) is the representation space
of $\chi$ and where $\Gamma$ acts on ${\frak M}\otimes F_{\chi}$
by the rule $\gamma\cdot(m,f)=(\gamma\cdot m,\chi(\gamma)f)$ for
$(\gamma,m,f)\in (\Gamma \otimes {\frak M}\otimes F_{\chi})$. Let
${\frak L}_{\Gamma}$ be the Laplace--Beltrami operator of $X$
acting on smooth sections of $V_{\chi}$; we obtain ${\frak
L}_{\Gamma}$ by projecting the Laplace--Beltrami operator of
${\frak M}$ (which is $G-$ invariant and thus $\Gamma-$ invariant)
to $X$.
%Let $a_0, n_0$
%denote the Lie algebras of $A, N$ in an Iwasawa decomposition $G=KAN$.
%Since the rank of $G$ is one,
%$\dim a_0=1$ by definition, say $a_0={\Bbb R}H_0$ for a suitable basis vector
%$H_0$. One can normalize the choice of $H_0$ by $\beta(H_0)=1$, where
%$\beta: a_0\rightarrow{\Bbb R}$ is the positive root which defines $n_0$;
%for more
%detail see Ref. \cite{will97-38-796}. Since $\Gamma$ is torsion free, each
%$%\gamma\in\Gamma-\{1\}$ can be represented uniquely as some power of a
%primitive
%element $\delta:\gamma=\delta^{j(\gamma)}$ where $j(\gamma)\geq 1$ is an
%integer and
%$\delta$ cannot be written as $\gamma_1^j$ for $\gamma_1\in \Gamma$, \,\,
%$j>1$ an
%integer. Taking $\gamma\in\Gamma$, $\gamma\neq 1$, one can find $t_\gamma>0$
%and $m_{\gamma}\in {\frak M} \stackrel{def}{=}\{m_{\gamma}\in K | m_{\gamma}a=
%am_{\gamma}, \forall a\in A\}$ such that $\gamma$
%is $G$ conjugate to $m_\gamma\exp(t_\gamma H_0)$, namely for some
%${\rm g}\in G, \,{\rm g}\gamma {\rm g}^{-1}=m_\gamma\exp(t_\gamma H_0)$.
%Besides let
%$\chi_{\sigma}(m) = {\rm trace}(\sigma(m))$ be the character of $\sigma$,
%for $\sigma$ a finite-dimensional representation of ${\frak M}$.
For any representation $\chi: \Gamma\rightarrow U(n)$ one can construct a
vector bundle ${\Bbb {\widetilde   E}}_{\chi}$ over a certain 4-manifold
$Y$ with
boundary $\partial Y=X$ which is an extension of a
flat vector bundle ${\Bbb E}_{\chi}$ over $X$.
Let ${\widetilde   A}_{\chi}$ be any extension of a flat connection
$A_{\chi}$
corresponding to $\chi$.

The index theorem of Atiyah--Patodi--Singer for the twisted Dirac
operator $D_{{\widetilde   A}_{\chi}}$
\cite{Atiyah75,Atiyah75a,Atiyah76}  gives here
$$
{\rm Index}\left(D_{{\widetilde A}_{\chi}}\right)=\int_{Y}{\rm ch}
({\Bbb {\widetilde E}}_{\chi})
{\widehat A}(Y)
-\frac{1}{2}(\eta(0,{\frak D}_{\chi})+
h(0,{\frak D}_{\chi}))
\mbox{,}
\eqno{(3.1)}
$$
where ${\rm ch}({\Bbb {\widetilde   E}}_{\chi})$ and ${\widehat
A}(Y)$ are the Chern character and ${\widehat  A}-$genus
respectively, ${\widehat  A}=1-p_1(Y)/24,\,p_1(Y)$ is the 1--st
Pontryagin class, $\eta(0, {\frak D_{\chi}})$ is the holomorphic
eta function, $h(0,{\frak D}_{\chi})$ is the dimension of the
space of harmonic spinors on $X$\, ($h(0,{\frak D}_{\chi}) ={\rm
dim} {\rm ker}\,{\frak D}_{\chi}$ = multiplicity of the
0--eigenvalue of ${\frak D}_{\chi}$ acting on $X$); and ${\frak
D}_{\chi}$ is a Dirac operator on $X$ acting on spinors with
coefficients in $\chi$. The Chern--Simons invariants of $X$ can be
derived from Eq. (3.1). Indeed we have:
$$
CS(A_{\chi})\equiv \frac{1}{2}\left({\rm dim}{\chi}\eta(0,{\frak
D})- \eta(0,{\frak D}_{\chi})\right), \,\,\,\,\,\,\,{\rm
mod}({\Bbb Z}/2) \mbox{.} \eqno{(3.2)}
$$

A remarkable formula relating $\eta(s,{\frak D})$, to the closed
geodesics on an hyperbolic manifold of $(4n-1)-$ dimension has
been derived in \cite{Millson,Moscovici}. More explicitly the
following function can be defined, initially for $\Re(s^2)\gg 0$,
by the formula
$$
{\rm log}{Z}_{\Gamma}(s)\stackrel{def}{=}
\sum_{[\gamma]\in {\mathcal  E}_1(\Gamma)}
(-1)^q\frac{L(\gamma,{\frak D})}
{|{\rm det}(I-P_h(\gamma))|^{1/2}}\frac{e^{-s\ell(\gamma)}}{m(\gamma)}
\,
\mbox{,}
\eqno{(3.3)}
$$
where ${\mathcal  E}_1(\Gamma)$ is the set of those conjugacy classes
$[\gamma]$
for which $X_{\gamma}$ has the property that the Euclidean de Rham factor
of ${\widetilde X}_{\gamma}$ is 1-dimensional (${\widetilde X}$ is a simply
connected cover of $X$ which is a symmetric space of noncompact type), the
number $q$ is half the
dimension of the fibre of the centre bundle $C(TX)$ over $X_{\gamma}$,
$\ell(\gamma)$ is the length of the closed geodesic
$c_{\gamma}$ (with multiplicity $m(\gamma)$) in the free homotopy class
corresponding to $[\gamma]$, $P_h(\gamma)$ is the restriction of the linear
Poincar\'{e} map $P(\gamma)$ at $(c_{\gamma},\dot{c}_{\gamma})\in TX$
to the directions normal to the geodesic flow and
$L(\gamma,{\frak D})$ is the Lefschetz number (see Ref. 16).

\subsection{The twisted Dirac operator}

Let now $\chi: \Gamma\rightarrow U(F)$ be a unitary representation
of $\Gamma$ on $F$. The Hermitian vector bundle ${\Bbb
F}={\widetilde X}\times_{\Gamma}F$ over $X$ inherits a flat
connection from the trivial connection on ${\widetilde X}\times
F$. We specialize to the case of locally homogeneous Dirac
operators ${\frak D}: C^{\infty}(X,{\Bbb E})\rightarrow
C^{\infty}(X,{\Bbb E})$ in order to construct a generalized
operator ${\mathcal  O}_{\chi}$, acting on spinors with
coefficients in $\chi$. If ${\frak D}: C^{\infty}(X,V)\rightarrow
C^{\infty}(X,V)$ is a differential operator acting on the sections
of the vector bundle $V$, then ${\frak D}$ extends canonically to
a differential operator ${\frak D}_{\chi}:
C^{\infty}(X,V\otimes{\Bbb F})\rightarrow
C^{\infty}(X,V\otimes{\Bbb F})$, uniquely characterized by the
property that ${\frak D}_{\chi}$ is locally isomorphic to ${\frak
D}\otimes...\otimes {\frak D}$\,\,\,(${\rm dim}\,F$ times)
\cite{Moscovici}. One can repeat the arguments to construct a
twisted zeta function ${Z}_{\Gamma}^{\chi}(s)$. It follows that
\cite{Bytsenko98,Bytsenko99,Bytsenko99a}: ${Z}_{\Gamma}^{\chi}(0)=
{Z}_{\Gamma}^{\chi}(0)^{{\rm dim}\,{\chi}} \exp\left(-2i\pi
CS(A_{\chi})\right)$, and eventually the Chern--Simons functional
takes the form
$$
CS(A_{\chi})\equiv \frac{1}{2i\pi}{\rm
log}\,\left[\frac{Z_{\Gamma}(0)^ {{\rm
dim}\,\chi}}{Z_{\Gamma}^{\chi}(0)}\right], \,\,\,\,\,\,\,\,\,{\rm
mod}({\Bbb Z}/2) \mbox{.} \eqno{(3.4)}
$$

\subsection{Hyperbolic manifolds with cusps}

We have used the formula for the relation between the eta
invariant of the signature operator and the Selberg zeta function
of odd type, which is defined by the geometric data for a compact
hyperbolic manifold. The corresponding formula between the
analytic torsion and the Ruelle zeta function for odd dimensional
hyperbolic manifold has been proven in Ref. [20]. In
addition, the symmetry that is similar to the Lefschetz fixed poin
theorem has been used to reduce the combination of the Selberg
zeta functions to the Ruelle zeta function. Later, these results
were generalized to the case of compact locally symmetric spaces
of higher ranks \cite{Moscovici,Moscovici1}, and they were
extended to the case of non--compact locally symmetric spaces with
finite volumes \cite{Park}. In the extension of these results
there are two main difficulties: the fact that the heat kernel of
the Laplace operator is not a trace class operator, and the
correct analysis of the weighted orbital integral term which
appears in the Selberg trace formula for a non--compact locally
symmetric space. These difficulties has been overcome in Ref.
[22] with the help of relative spectral invariants
\cite{Muller} and the Fourier transform of the weighted orbital
integral for the ${\Bbb R}-$ rank one cases \cite{Hoffmann},
respectively.

\subsection{Structure on symmetric spaces}
Let us explain the results more precisely. Consider $G$, a
noncompact semisimple Lie group with finite center $Z_G$, $K$ its
maximal compact subgroup, and $G/K$ to be a noncompact symmetric
space of ${\Bbb R}-$rank one. Let $\Gamma$ a torsion free discrete
subgroup of $G$ with the volume of $\Gamma\backslash G$ finite.
Taking into account the fixed Iwasawa decomposition $G=KAN$,
consider a $\Gamma-$cuspidal minimal parabolic subgroup $P$ of $G$
with the Langlands decomposition $P=BAN$, being $B$ the
centralizer of $A$ in $K$.
Let us define the Dirac operator ${\frak D}$, assuming a spin
structure for $X = \Gamma \backslash ({\rm Spin}(2n+1, 1)/{\rm
Spin}(2n+1))$. The spin bundle $E_{\tau_s}$ is the locally
homogeneous vector bundle defined by the spin representation
$\tau_s$ of the maximal compact group ${\rm Spin}(2n+1)$. One can
decompose the space of sections of $E_{\tau_s}$ into two
subspaces, which are given by the half spin representations
$\sigma_{\pm}$ of ${\rm Spin}(2n)\subset {\rm Spin}(2n+1)$.

The Selberg trace formula could be used to prove a relation of the
eta invariant and the Selberg zeta function of  odd type. Let us
consider a family of functions ${\mathcal K}_t$ over $G= {\rm
Spin}(2n+1, 1)$, which is given by taking the local trace for the
integral kernel $\exp(-t{\frak D}^2)$ (or ${\frak D}\exp(-t{\frak
D}^2)$). The Selberg trace formula applied to the scalar kernel
${\mathcal K}_t$ has the form \cite{Park}:

$$
\!\!\!\!\!\!\!\!\!\!\!\!\!\!\!\!\!\!\!\!\!\!\!\!\!\!\!\!\!\!
\!\!\!\!\!\!\!\!\!\!\!\!\!\!\!\!\!\!\!\!\!\!\!\!\!\!\!\!\!\!
\!\!\!\!\!\!\!\!\!
\sum_{\sigma = \sigma_{\pm}}\sum_{\lambda_k\in \sigma_p}
{\hat {\mathcal K}}_t(\sigma, i\lambda_k)
-\frac{d(\sigma_{\pm})}{4\pi}\int_{\Bbb R}
{\rm Tr}\left(S_{\Gamma}(\sigma_{\pm},-i\lambda)
\right.
$$
$$
\left.
\times
(d/ds)S_{\Gamma}(\sigma_{\pm},s)|_{s=i\lambda}\pi_{\Gamma}
(\sigma_{\pm},i\lambda)
({\mathcal K}_t)\right)d\lambda
= I_{\Gamma}({\mathcal K}_t) + H_{\Gamma}({\mathcal K}_t)
+ U_{\Gamma}({\mathcal K}_t)\,
\mbox{,}
\eqno{(3.5)}
$$
where $\sigma_p \in {\rm Spec}\,{\frak D}$, $d(\sigma_{\pm})$ is the
degree of the half spin representation of ${\rm Spin}(2n)$,
$S_{\Gamma}(\sigma_{\pm}, i\lambda)$ is the scattering matrix and
$I_{\Gamma}({\mathcal K}_t),\, H_{\Gamma}({\mathcal K}_t)$ and
$U_{\Gamma}({\mathcal K}_t)$ are identity, hyperbolic and
unipotent orbital integrals respectively. The analysis of the
unipotent orbital integral $U_{\Gamma}({\mathcal K}_t)$ gives the
following result \cite{Park}: all the unipotent terms are
vanishing in the Selberg trace formula applied to the odd kernel
function. This means that we can get the same formula as the
result given in Ref. [15] in the relation of the eta
invariant and the Selberg zeta function of odd type for the
compact hyperbolic manifolds without any additional terms, which
might exist due to the cusps.
\\

{\sf Note.} Observe \cite{Park} that in the functional
equation of the eta invariant and the Selberg zeta function of odd
type, a term ${\rm tr}_s(S_{\Gamma}(\sigma_{\pm},
-z)(d/dz)S_{\Gamma} (\sigma_{\pm},z))$, given by the scattering matrix
$S_{\Gamma}(\sigma_{\pm}, z)$ appears, where the supertrace ${\rm
tr}_s$ is the trace taken over the subspace related to the
representation space of $\sigma_{\pm}$ with weight $\pm 1$. This is an
odd type formula, for the functional equation of Selberg zeta
functions, which has been proved in Ref. [25].

%\vskip .55cm
%\\
%\\

\section{Results on the Holographic Principle}

According to the holographic principle, there exist strong ties
between certain field theories on a manifold (``bulk space'') and
on its boundary (at infinity). A few mathematically exact results
relevant to that program are the following. The class of Euclidean
AdS$_3$ spaces which we have considered here are quotients of the
real hyperbolic space ${\Bbb H}^3$ by a Schottky group. The
boundaries of these spaces can be compact oriented surfaces with
conformal structure (compact complex algebraic curves).

In Ref. [26], a principle associated with the Euclidean
AdS$_2$ holography has been established. The bulk space is there a
modular curve, which is the global quotient of the hyperbolic
plane ${\Bbb H}^2$ by a finite index group, $\Gamma$, of $G =
PSL(2, {\Bbb Z})$. The boundary at infinity is then ${\Bbb
P}^1({\Bbb R})$. Let ${\Bbb P}$ be a coset space ${\Bbb P} =
G/\Gamma$. Then, the modular curve $X_G:= \Gamma\backslash {\Bbb
H}^2$ can be presented as the quotient $X_G = G\backslash({\Bbb
H}^2\times {\Bbb P})$; its non--commutative boundary (in the sense
of Connes \cite{Connes}) as the $C^*-$ algebra $C({\Bbb P}^1({\Bbb
R})\times {\Bbb P}){>\!\!\!\lhd} G$, Morita equivalent to $C({\Bbb
P}^1({\Bbb R})){>\!\!\!\lhd}\Gamma$ \cite{ManMar,Mar,Manin}. The
results which have been regarded as manifestations of the
holography principle are \cite{Manin}:

\vskip .25cm
\begin{enumerate}
\item There is a correspondence between the eigenfunctions of the
transfer operator $L_s$ and the eigenfunctions of the Laplacian
(Maas wave forms). This correspondence is established in Refs.
[30,31,32]. \item The cohomology classes in
$H_1(X_G, {\rm cusps}, {\Bbb R})$ can be regarded as elements in
the cyclic cohomology of the algebra $C({\Bbb P}^1({\Bbb R})\times
{\Bbb P}){>\!\!\!\lhd} G$. Cohomology classes of certain geodesics
in the bulk space correspond to projectors in the algebra of
observables on the boundary space. \item An explicit
correspondence exists between a certain class of fields in the
bulk space (Mellin transforms of modular forms of weight two) and
the class of fields on the boundary.
\end{enumerate}
%\vskip .55cm
In the three--dimensional case we can interprete the statement in
Eq. (2.6) as an instance of a kind of holographic correspondence,
if we regard ${\rm det} {\frak L}_{\Gamma}$ as a partition
function in the bulk space and the right hand side as a theory
related to the boundary at infinity.

Other constructions associated with the symmetric space can be
considered for convex--cocompact groups. In fact, let $\partial X$
be a geodesic boundary of the symmetric space $X$ of a real, rank
one, semisimple Lie group $G$. If $\Gamma \subset G$ is a discrete
torsion--free subgroup, then a $\Gamma-$equivalent decomposition,
$\partial X =\Omega \cup \Lambda$, can be constructed, where
$\Lambda$ is the limit set of $\Gamma$. The subgroup $\Gamma$ is
called convex--cocompact if $\Gamma\backslash X\cup \Omega$ is a
compact manifold with boundary \cite{Bunke}. The orbit space,
$X=\Gamma\backslash {\Bbb H}^{n}$, may be viewed as the interior
of a compact manifold with boundary, namely the Klein manifold for
$\Gamma$ \cite{Patterson}, ${\overline X} = (\Gamma\backslash
{\Bbb H}^{n})\cup (\Gamma\backslash \Omega(\Gamma))$, so that the
boundary at infinity is given by $\partial_{\infty}X =
\partial {\overline{X}} = \Gamma\backslash \Omega(\Gamma)$.

\end{document}